\documentclass[epj]{svjour}

\usepackage{graphicx}
\usepackage{dcolumn}
\usepackage{bm}
\usepackage{graphics}
\usepackage{epsfig,color}

\newcommand{\be}{\begin{equation}}
\newcommand{\ee}{\end{equation}}
\newcommand{\bea}{\begin{eqnarray}}
\newcommand{\eea}{\end{eqnarray}}
\newcommand{\la}{\langle}
\newcommand{\ra}{\rangle}

\begin{document}
\title{Triplet superconductivity in a 1D itinerant electron system with 
transverse spin anisotropy}
\titlerunning{Triplet superconductivity in a 1D itinerant electron system}
\author{C. Dziurzik$^{1}$, G.I. Japaridze$^{2}$\thanks{Permanent address: 
Andronikashvili Institute of Physics, Georgian Academy of Sciences, 
Tamarashvili 6, Tbilisi 0177, Tbilisi, Georgia},
A. Schadschneider$^{1}$, I. Titvinidze$^{3a}$, J. Zittartz$^{1}$} 

\authorrunning{C.\ Dziurzik et al.}
%
%
\institute{$^{1}$ Institut f\"ur Theoretische Physik, Universit\"at zu K\"oln, 
D-50937 K\"oln, Germany\\
$^{2}$ Max-Planck-Institut f\"ur Physik komplexer Systeme, N\"othnitzer
Str. 38, D-01187 Dresden, Germany \\
$^{3}$ Institut f\"ur Theorie der Kondensierten Materie Universit\"at 
Karlsruhe, D-76128 Karlsruhe, Germany}

\date{Received: \today }

\abstract{In this paper we study the ground state phase diagram of a 
one-dimensional $t-J-U$ model away from half-filling. In the large-bandwidth 
limit and for ferromagnetic exchange with easy-plane anisotropy a phase 
with gapless charge and massive spin excitations, characterized by the 
coexistence of triplet superconducting and spin density wave 
instabilities is realized in the ground state. 
With increasing ferromagnetic exchange transitions into a ferrometallic
and then a spin gapped triplet superconducting phase take place.}   

\PACS{
{71.10.Hf}{}
\and 
{71.10.Fd}{}
\and
{74.20.Mn}{}
\and
{71.27.+a}{}
\and 
{75.10.Pq}{}
}

\maketitle

\section{Introduction}\label{intro}

Experimental and theoretical investigations show
that many strongly correlated electronic systems
exhibit surprisingly complex phase diagrams \cite{dagotto05}.
Especially the competition or coexistence of magnetic order and 
superconductivity is a topic of increased current interest.
Magnetically mediated Cooper pairing near the antiferromagnetic
instability is widely discussed in the context of superconductivity in
copper-oxide systems \cite{Chubukov_Pines_book_2003,Manske_book_2004}.
Moreover, the discovery of Triplet Superconductivity (TS) in
$Sr_{2}RuO_{4}$ \cite{MacKenzieMaeno} and the recent discovery of
coexistence of the TS phase with ferromagnetism in $UGe_{2}$
\cite{UGE2}, $URhGe$ \cite{URhGe} and $ZrZn_2$ \cite{ZrZn2} has
triggered an increased activity in studies of correlated electron
models showing close proximity of triplet superconducting and
ferromagnetically ordered phases \cite{RS,MS,Sigrist,Kirpatrick1,Singh_Mazin_2002,WalkerSamokhin,Chubukov2,Buzdin,Kirpatrick2,Mazin_Singh_2004}.

Another challenging problem is superconductivity in quasi-one-dimensional 
compounds \cite{Ishiguro}. More than two decades
have passed since the discovery of superconductivity in so-called
Bechgaard salts such as (TMTSF)$_2$X with X=PF$_6$, ClO$_4$, etc.
\cite{Jerome_1980}.  Most interesting is the phase diagram of
$(TMTSF)_{2}PF_{6}$ which shows a spin-Peierls (SP) phase in the
ground state at atmospheric pressure. Increasing pressure leads first
to a transition from the SP phase into a spin density wave (SDW)
phase, and finally to the suppression of the SDW ground state in favor
of superconductivity \cite{Jerome_1991}. In last years an increasing
amount of convincing experimental evidence has been accumulated in
favour of the triplet superconducting ordering in these compounds
\cite{TrSc_BechgaardSalts_Exp}.  This triggered interest in
low-dimensional models of correlated electrons showing mechanism for
Cooper pairing of triplet symmetry coexisting or closely competing
with SDW type ordering.  Although a wide variety of theoretical
approaches have so far been used to study this topic
\cite{Lebed_2000,Kuroki_2001,Siginishi_2004,Tanaka_Kuroki_2004,Kuroki_et_al_2004,FS_CM_0411013,Nishimoto_et_al_2005},
the understanding of the microscopic mechanisms for the complex phenomena
in these compounds is still far from complete.

In \cite{JM_2000} an extension of the Hubbard model including
anisotropic spin-spin interactions has been proposed as a suitable
model for systems with coexisting orders. The model describes a system
of itinerant electrons with anisotropic spin-exchange interaction
between electrons on nearest-neighbor sites.  The one-dimensional
version of the Hamiltonian reads:
\begin{eqnarray}\label{itinerant-t-U-J-model}
  {\cal H} & = &  -t\sum_{n,\alpha}(c^{\dagger}_{n,\alpha}
  c^{\vphantom{\dagger}}_{n+1,\alpha}
  +c^{\dagger}_{n+1,\alpha}c^{\vphantom{\dagger}}_{n,\alpha}) + 
  U \sum_{n}\rho_{n,\uparrow}\rho_{n,\downarrow}\nonumber\\
  &+&\frac{J_{xy}}{2}\sum_{n}(S^{+}_{n}S^{-}_{n+1} + h.c.)
  + J_{z}\sum_{n}S^{z}_{n}S^{z}_{n+1}\,.
\end{eqnarray}
Here $c^{\dagger}_{n,\alpha}$ ($c^{\vphantom{\dagger}}_{n,\alpha}$) is
the creation (annihilation) operator for an electron at site $n$ with spin
${\alpha}$, $\rho_{,\alpha}(n)=c^{\dagger}_{n,\alpha}
c^{\vphantom{\dagger}}_{n,\alpha}$, $\vec{S}(n)=\frac{1}{2}
c^{\dagger}_{n,\alpha}\vec{\sigma}^{\vphantom{\dagger}}_{\alpha\beta}
c^{\vphantom{\dagger}}_{n,\beta}$
where $\sigma^{{\it i}}$ (${\it i}=x,y,z$) are the Pauli matrices.

Indeed, this model was shown to exhibit an extremely rich phase 
diagram in the case of a half-filled band \cite{JM_2000,DJSZ_2004}. 
In particular in the case of the ferromagnetic $XY$-type exchange
interactions the ground state phase diagram consists of sequence of
transitions (with increasing ferro exchange) from a metallic phase with 
coexisting TS$^{0}$ and SDW$^{z}$ instabilities, into an insulating N\'eel type
antiferromagnetic phase and finally, for strong $XY$-type ferro-exchange 
into the insulating \emph{ferromagnetic} $XY$ phase \cite{DJSZ_2004}.

In this paper we study the {\it effect of doping} on the ground state
phase diagram of the model (\ref{itinerant-t-U-J-model}). We use a
combined approach based on continuum-limit bosonization 
and DMRG techniques. In the range of applicability of the
continuum-limit approach we have obtained the ground state phase
diagram for a wide range of model parameters and band-fillings $\nu$.  In our
numerical studies we restrict our consideration to the case of a
quarter-filled band with $XY$-type spin exchange interaction and
on-site repulsion $U\geq 0$. We investigate the excitation spectrum of the
system as well as the behavior of various correlation functions.
Depending on the relation between the model parameters $J_{xy}/t$ and
$U/t$ we have shown the existence of four different metallic phases in the
ground state (see Fig.~\ref{Fig:TRSC3_Fig_10}): In the case of
antiferromagnetic exchange the system shows properties of a spin gapped
(Luther-Emery) metal with coexistence of the singlet-superconducting
and charge-density-wave (CDW) instabilities. The line $J_{xy}=0$
corresponds to the Luttinger Liquid phase and marks the transition
from a spin gapped sector with singlet Cooper pairing (antiferromagnetic
exchange) into the spin gapped sector with triplet Cooper pairing
(ferromagnetic exchange). For weak ferromagnetic exchange, 
at $J^{c1}_{xy} < J_{xy}<0$
the system displays properties of a spin gapped metal with coexistence
of the triplet superconducting and spin-density-wave (SDW$^{z}$)
instabilities. For ferromagnetic spin coupling the spin gap
dependence on strength of the transverse exchange exhibits a
dome-type shape, opening at $J_{xy}=0$ and closing at
$J_{xy}=J^{c1}_{xy}$. At $J_{xy}<J^{c1}_{xy}$ a rather unconventional
ferrometallic phase with gapless spin excitations and strongly
dominating transverse ferromagnetic and triplet superconducting
instabilities in the ground state. Finally, at $J_{xy}<J^{c2}_{xy}$,
our numerical data indicates on opening of the spin gap. In this phase
the system is expected to show properties of a triplet superconductor.

The paper is organized as follows: in the next section the
weak-coupling continuum-limit version of the model is investigated. 
This allows to derive the weak-coupling phase diagram 
(Sec.~\ref{sec-weak}).
In Sec.~\ref{sec:NumericalResults} results of DMRG studies 
for chains up to $L=120$ sites are presented. 
Finally, Sec.~\ref{sec-conclusion} is devoted to a discussion and 
concluding remarks.


\section{\bf Continuum-Limit Bosonization}
\label{sec-continuum}

Below in this section we consider the low-energy effective field
theory of the model (\ref{itinerant-t-U-J-model}) in the case of
non-half-filled band.

The standard bosonization procedure allows to express the initial lattice
model in terms of two independent bosonic Hamiltonians 
$$
{\cal H} = {\cal H}_{c} + {\cal H}_{s}
$$
describing respectively the {\em charge} ($c$) and {\em spin} ($s$) degrees of
freedom. For the band-filling $\nu \neq 1/2$, the  gapless charge
sector is described by the free Bose field Hamiltonian
\begin{equation}\label{KGc}
H_{c} = \frac{v_{c}}{2} \int dx\Big\{\,\frac{1}{K_{c}}(\partial_{x}
\varphi_{c})^2 + K_{c} (\partial_x \vartheta_{c})^2 \Big\},
\end{equation}
while the spin sector is governed by the quantum Sine-Gordon field
\begin{eqnarray}
&&H_{s} = \frac{v_{s}}{2}
\int dx\Big\{\, \frac{1}{K_{s}}(\partial_{x}\varphi_{s})^2
+ K_{s} (\partial_x\vartheta_{s})^2\nonumber\\
&&\qquad\qquad\qquad\qquad 
+\frac{2m_{s}}{a_0^2}\cos(\sqrt{8\pi}\varphi_{s})\Big\}\, .
\label{SGs}
\end{eqnarray}
Here $\varphi_{c,s}(x)$ and $\vartheta_{c,s}(x)$ are mutually dual 
bosonic fields
\bea
&[\varphi_{c,s}(x),\vartheta_{c,s}(x)]={i\over 2},&\,\, \nonumber\\
&\partial_t\varphi_{c,s}= v_{c,(s)} \partial_x\vartheta_{c,s} \,, \,\, 
 \partial_x\varphi_{c,s}={1\over v_{c,(s)}}\partial_t\vartheta_{c,s}\, ,&
\eea 
and $a_{0}$ is the infrared cutoff of the theory. The model parameters are given by
\bea
2(K_{c}-1) &\equiv& g_{0c}\nonumber\\ 
&=& - \frac{1}{\pi v_{F}}\Big[U - ( J_{xy} + \frac{1}{2} J_{z} )
  \cos(2\pi\nu)\Big],\\
2(K_{s}-1)  &\equiv&  g_{0s}\nonumber\\  
&=&\frac{1}{\pi v_{F}} \Big[U-J_{z}-(J_{xy}-\frac{J_{z}}{2} )
  \cos(2\pi\nu)\Big],\label{Ks} \\
 2\pi m_{s}  &\equiv& g_{\perp}\nonumber\\ 
&=& \frac{1}{\pi v_F}\Big[U-J_{xy}-\frac{J_{z}}{2}\cos(2\pi\nu)\Big]
\label{Ms}
\eea
and the velocities of charge and spin excitations 
$v_{c,(s)} = v_{F}/K_{c,(s)}$, where $v_{F}=2t a_{0} \sin(\pi \nu)$.

Since at $\nu \neq 1/2$ the charge sector is described by the free 
Gaussian field (\ref{KGc}) the vacuum averages of exponentials of the charge 
fields show a power-law decay at large distances 
\begin{eqnarray}
  \langle 
  e^{i \sqrt{2\pi} \varphi_{c}(x)} 
  e^{-i \sqrt{2\pi} \varphi_{c}(x')}
  \rangle 
  &\sim& \left| x - x' \right|^{-K_{c}},
  \label{freecorrelations1}\\
  \langle 
  e^{i \sqrt{2\pi} \vartheta_{c}(x)} 
  e^{-i \sqrt{2\pi} \vartheta_{c}(x')}
  \rangle
  &\sim& \left| x - x' \right|^{-1/K_{c}},
  \label{freecorrelations2}
\end{eqnarray}
and the only parameter controlling contribution of the gapless charge degrees 
of freedom to the infrared properties of the system is the charge LL parameter 
$K_{c}$. 

The infrared behavior of the Sine-Gordon Hamiltonian ${\cal H}_{s}$ is
described by the corresponding pair of renormalization group (RG)
equations for the effective coupling constants $K_{s}(l)$ and
$M_{s}(l)$
\begin{eqnarray}\label{RGeq}
\frac{dM_{s}(L)}{dL} &=&-2(K_{s}(L)-1)M_{s}(L)\nonumber\\
\frac{dK_{s}(L)}{dL}&=&-\frac{1}{2}M_{s}^{2}(L)
\end{eqnarray}
where $L=\ln(a_{0})$, $K_{s}(L=0) = 1 + \frac{1}{2}g_{0s}$ and 
$M_{s}(L=0)=g_{\perp}/2\pi$. The pair of RG equations (\ref{RGeq}) describes 
the Kosterlitz-Thouless transition \cite{KT} in the spin channel. The flow 
lines lie on the hyperbola
\be
4(K_{s}-1)^{2}-M_{s}^{2}=\mu_{\pm}^{2}=g_{0s}^{2}-g_{\perp}^{2}
\label{RGeq2}
\ee
and depending on the relation between the bare coupling constants 
$g_{0s}$ and $g_{\perp}$ exhibit two different regimes \cite{Wiegmann}:

{\em Weak-coupling regime.} \ \ For $g_{0s}\geq \left|g_{\perp} \right|$
we are in the weak-coupling regime with effective mass 
${\cal M}_{s} \to 0$. The low energy (large distance) behavior of the 
corresponding gapless mode is described by a free scalar field.

The vacuum averages of exponentials of the corresponding fields show
a power-law decay at large distances,
\begin{eqnarray}
  \langle 
  e^{i \sqrt{2\pi} \varphi_{s}(x)} 
  e^{-i \sqrt{2\pi} \varphi_{s}(x')}
  \rangle 
  &\sim& \left| x - x' \right|^{- K^{\ast}_{s}},
  \label{freecorrelations1a}\\
  \langle 
  e^{i \sqrt{2\pi} \vartheta_{s}(x)} 
  e^{-i \sqrt{2\pi} \vartheta_{s}(x')}
  \rangle
  &\sim& \left| x - x' \right|^{-1/K^{\ast}_{s}},
  \label{freecorrelations2a}
\end{eqnarray}
and the only parameter controlling the infrared behavior in the gapless
regime is the fixed-point value of the effective coupling constants
$K^{\ast}_{s} = K_{s}(l=\infty) $ determined from 
Eq.~(\ref{RGeq2}).

{\em Strong coupling regime.} \ \ For $2(K^{0}_{s}-1) <
\left|M^{0}_{s}\right|$ the
system scales to strong coupling: depending on the sign of the bare
mass $M^{0}_{s}$, the renormalized mass ${\cal M}_{s}$ is driven to
$\pm\infty$,
signaling a crossover to one of two strong coupling regimes with a
dynamical generation of a commensurability gap in the excitation
spectrum. The flow of $\left|{\cal M}_{s}\right|$ to large values indicates
that the ${\cal M}_{s}\mbox{cos}\sqrt{8\pi}\phi_{s} $ term in the
sine-Gordon model dominates the long-distance properties of the
system. Depending on the sign of the mass term, the field $\varphi_{s}$ gets
ordered with expectation values \cite{ME}
\begin{equation}
  \label{orderfields}
  \langle\varphi_{s}\rangle =\left\{ \begin{array}{l@{\quad}}
      \sqrt{\pi/8}
      \hskip0.5cm(M^{0}_{s}>0) \\
   \hskip0.50cm   0 \hskip0.72cm (M^{0}_{s}<0)\end{array}\right. \,.
\end{equation}

It easy to check that, using the initial values of the coupling
constants given in (\ref{Ks})-(\ref{Ms}), we obtain the following
condition for generation of a gap in the spin excitation spectrum
\bea
&&\left|U-J_{xy}-\frac{J_{z}}{2}\cos(2\pi\nu)\right|>\nonumber\\
&&\qquad U-J_{xy}\cos(2\pi\nu) - J_{z} + \frac{J_{z}}{2}\cos(2\pi\nu).
\eea

\subsection{Order parameters}

To clarify the symmetry properties of the ground states of the system in 
different sectors we consider the following set of order parameters 
corresponding to the smooth ``{\it sm}'' and staggered ``{\it st}'' 
parts of : 
\begin{itemize}
\item[1)] the on-site density operator 
\end{itemize}
$$
\rho(n) \Rightarrow \rho_{sm}(x) + \rho_{st}(x)
$$
where 
\begin{eqnarray}\label{Density}
\rho_{sm}(x)&\simeq& \sqrt{\frac{2}{\pi}}\partial_{x}\varphi_{c}(x),\\
\rho_{st}(x) &\equiv& {\cal O}_{CDW}(x) \nonumber\\
&\simeq&   \sin(\sqrt{2\pi}\varphi_{c} - 2k_{F}x) 
\cos( \sqrt{2\pi}\varphi_{s}),
\end{eqnarray}
\begin{itemize}
\item[2)] the on-site spin-density
\end{itemize}
\be
{\bf S}_{n} \Rightarrow {\bf S}_{sm}(x) + {\bf S}_{st}(x)
\ee
where
\begin{eqnarray}\label{Test}
 S^{z}_{sm}(n) &\equiv& {\cal O}^{z}_{FM}(x) \simeq  
 \frac{1}{\sqrt{2\pi}}\partial_{x}\varphi_{s}(x), \\
 S^{x}_{sm}(n) &\equiv& {\cal O}^{x}_{FM}(x)\nonumber\\  
& \simeq & \sin(\sqrt{2\pi}\varphi_{s})
\cos\left(\sqrt{2\pi}\vartheta_{s}\right),\\
 S^{y}_{sm}(n)&\equiv& {\cal O}^{y}_{FM}(x)\nonumber\\   
&\simeq &  \sin(\sqrt{2\pi}\varphi_{s})
\sin\left(\sqrt{2\pi} \vartheta_{s}\right),
\end{eqnarray}  
and
\begin{eqnarray}
 S^{z}_{st}(n) &\equiv&  {\cal O}_{SDW^{z}}(x)  \nonumber\\  
& \simeq&  \cos(\sqrt{2\pi}\varphi_{c}+ 2k_{F}x) 
\sin\left(\sqrt{2\pi}\varphi_{s}\right),\\
 S^{x}_{st}(n)& \equiv & {\cal O}_{SDW^{x}}(x) \nonumber\\      
&\simeq &  \cos(\sqrt{2\pi}\varphi_{c} + 2k_{F}x) 
\sin\left( \sqrt{2\pi}\vartheta_{s}\right),
\label{Sx}\\
 S^{z}_{st}(n) &\equiv&  {\cal O}_{SDW^{y}}(x)\nonumber\\     
& \simeq &   \cos(\sqrt{2\pi}\varphi_{c} +2k_{F}x )
\cos\left(\sqrt{2\pi}\vartheta_{s}\right).\label{Sy}
\end{eqnarray}  
In addition we use the following set of superconducting order parameters:
\begin{itemize}
\item[3a)] the on-site singlet
\end{itemize}
\be\label{SS} 
c^{\dagger}_{n,\uparrow}c^{\dagger}_{n,\downarrow}\Rightarrow 
{\cal O}^{\dagger}_{SS}(x) +  {\cal O}^{\dagger}_{\eta-SS}(x) 
\ee
\begin{itemize}
\item[3b)] the extended singlet
\end{itemize}
\bea\label{ES} 
\frac{1}{\sqrt{2}}\left(c_{n,\uparrow}^{\dagger}c_{n+1,\downarrow}^{\dagger} 
-  c_{n,\downarrow}^{\dagger}c_{n+1,\uparrow}^{\dagger}\right) 
\Rightarrow {\cal O}^{\dagger}_{ES}(x) + {\cal O}^{\dagger}_{\eta-ES}(x)
&&\nonumber\\ 
\eea
\begin{itemize}
\item[3c)] and the triplet pairing
\end{itemize}
\begin{eqnarray}
 \frac{1}{\sqrt{2}}  
\left(c_{n,\uparrow}^{\dagger}c_{n+1,\downarrow}^{\dagger} +  
c_{n,\downarrow}^{\dagger}c_{n+1,\uparrow}^{\dagger}\right) &\Rightarrow&
{\cal O}_{TS^{0}}^{\dagger}(x) +  {\cal O}_{\eta-TS^{0}}(x)\nonumber\\  
&&\label{TS-0}\\
\frac{1}{\sqrt{2}}
\left(c_{n,\uparrow}^{\dagger}c_{n+1,\uparrow}^{\dagger} \pm
c_{n,\downarrow}^{\dagger}c_{n+1,\downarrow}^{\dagger}\right)
&\Rightarrow & {\cal O}^\dagger_{TS^{xy}}(x)\, . 
\label{TS-xy}
\eea
Here operators without and with the subscipt $\eta$ correspond,
respectively, to the smooth and $2k_{F}$-modulated (staggered) parts
of the corresponding superconducting order parameters.

The bosonized expressions for the smooth parts of the corresponding 
superconducting order parameters, up to the accuracy of irrelevant 
$\nu$-dependent amplitudes and phase shifts are given by
\begin{eqnarray}
{\cal O}_{SS}^{\dagger}(x)  & \sim &   {\cal O}_{ES}^{\dagger}(x)
 \sim   \cos(\sqrt{2 \pi }\varphi_{s})\,
e^{{\it i} \sqrt{2\pi}\vartheta_{c}}\, ,
\label{SS-bosonized} \\
{\cal O}_{TS^{0}}^{\dagger}(x) &\sim &
  \sin(\sqrt{2\pi}\varphi_{s})\,
e^{{\it i} \sqrt{2\pi}\vartheta_{c}}\, , 
\label{TS-0-bosonized} \\ 
{\cal O}^\dagger_{TS^{xy}}(x) 
& \sim &
\left\{ \begin{array}{l@{\quad}}
\cos\big(\sqrt{2\pi}\vartheta_{s}\big) 
e^{{\it i}\sqrt{2\pi}\vartheta_{c}} \\
\sin\big(\sqrt{2\pi}\vartheta_{s}\big)
e^{{\it i}\sqrt{2\pi}\vartheta_{c}}
\end{array}\right. . 
\end{eqnarray}
Similarly the bosonized expression for the staggered components 
of the corresponding superconducting order parameters are given by
\begin{eqnarray}
&{\cal O}_{\eta-SS}^{\dagger}(x)  \sim  {\cal O}_{\eta-ES}^{\dagger}(x) 
\sim {\cal O}_{\eta-TS^{0}}^{\dagger}(x)& \nonumber\\  
& \sim   \sin(\sqrt{2\pi}\varphi_{c}+ 2k_{F}x)\, 
e^{{\it i} \sqrt{2\pi}\vartheta_{c}}\, ,&
\end{eqnarray}

Note that the smooth part in Eq.~(\ref{SS}) corresponds
to the usual BCS-type pairing while at half-filling the oscillating terms in
(\ref{SS}) and (\ref{ES}) describe the weak-coupling analogs of the 
$\eta$-pairing superconductivity \cite{Yang}.

\section{The Weak-Coupling Phase Diagram}
\label{sec-weak}

In this section we consider the ground state phase diagram of the model 
(\ref{itinerant-t-U-J-model}) away from the commensurate value of the 
1/2-filled band. Due to the invariance of the model parameters under the
transformation $\nu \rightarrow 1-\nu$ we restrict our consideration to the
sector $0<\nu<1/2$. Below we consider the the weak-coupling ground state 
phase at quarter-filling  in detail, while for $\nu \neq 1/4$ we present 
a qualitative description of the phase diagram. 

\subsection{The $U=0$ case.}

Let us start with the case $U=0$ where the basic equations read:
\begin{eqnarray}\label{Kc-U-0}
K_{c} \simeq  1 + \frac{1}{2\pi v_{F}}\left(J_{xy} + \frac{1}{2} J_{z}
\right)\cos(2\pi\nu)\, ,
\end{eqnarray}
and 
\bea
& 2(K_{s}-1)  \simeq -\frac{1}{\pi v_{F}} \Big[J_{z}+(J_{xy} 
- \frac{1}{2} J_{z})\cos(2\pi\nu)\Big]\, ,&\label{Ks-U-0}\\
&2\pi m_{s} \simeq -\frac{1}{\pi v_F}\Big[J_{xy} 
+ \frac{1}{2}J_{z}\cos(2\pi\nu)\Big]\, .&
\label{Ms-U-0}
\eea

\begin{figure}[tbh]
\begin{center}
\includegraphics[width=0.4\textwidth]{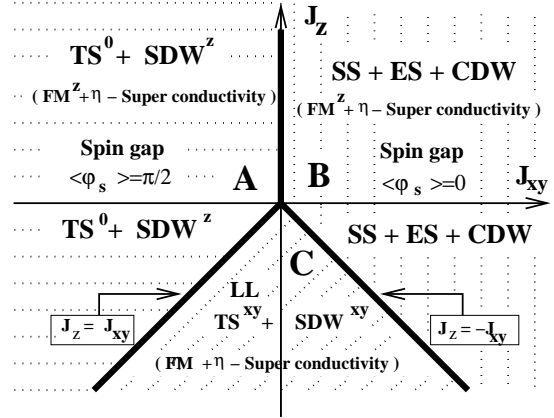}
\caption[DIAGONAL]{The ground state phase diagram of
  the 1/4-filled itinerant $t-J_{xy}-J_z$ model. 
  The thick solid lines indicate borders between the weak-coupling limit
  phases. Along these lines and in the sector C the spin excitations 
  are gapless. The charge excitation spectrum is gapless in all sectors
  of the phase diagram.}
\label{Fig:Fig-1_PD_025Filled_JxyJz.eps}
\end{center}
\end{figure}

In the following we study the two cases $\nu = 1/4$ and $\nu \neq 1/4$
separately.

\subsubsection{$\nu = 1/4$}
\label{sec-u0nu4}

At $U=0$ the charge sector is featureless with $K_{c}=1$ independently
from the values of the exchange couplings $J_{z}$ and $J_{xy}$. Since
the charge sector is featureless, the ground state phase diagram is
completely determined only by the spin degrees of freedom. The
condition for dynamical generation of a spin gap reads
$\left|-J_{xy}\right| > - J_{z}$ and thus the spin sector is gapless
for $J_{z} \leq -\left|J_{xy}\right|$ and along the
semi-axis $J_{z}\geq 0$.  These conditions determine the following
three sectors of the phase diagram (see Fig.
\ref{Fig:Fig-1_PD_025Filled_JxyJz.eps}):

The sector {\bf A} ($J_{xy}<0, J_{z}>-J_{xy}$) corresponds to a spin gapped
phase with dominating $SDW^{z}$ and $TS^{0}$ ordering. These correlations 
decay as power-laws at large distances:
\bea
\langle {\cal O}_{SDW^{z}}(0) {\cal O}_{SDW^{z}}(r)\rangle &\simeq &
\cos(\pi r/2)\cdot r^{-1} \, , 
 \label{JxyJz_A_SDWz}\\
\langle {\cal O}_{TS^{0}}(0){\cal O}_{TS^{0}}(r)\rangle &\simeq& 
r^{-1}  \, . 
\label{JxyJz_A_TSz}
\eea
The longitudinal ferromagnetic and the $\eta$-superconducting correlations 
decay faster as 
\bea
&&\langle {\cal O}_{FM^{z}}(0){\cal O}_{FM^{z}}(r)\rangle \simeq  
r^{-2}\, , 
\label{FM-zz}\\ 
&&\langle {\cal O}_{\eta-SS}(0){\cal O}_{\eta-SS}(r)\rangle \simeq  
\langle {\cal O}_{\eta-ES}(0){\cal O}_{\eta-ES}(r)\rangle\nonumber\\
&&\quad\simeq \langle {\cal O}_{\eta-TS^{0}}(0){\cal O}_{\eta-TS^{0}}(r)
\rangle \simeq  \cos(\pi r/2)\cdot r^{-2} \, .
\label{eta}
\eea
Other correlations decay exponentially. 

The sector {\bf B} ($J_{xy}>0, J_{z}>-J_{xy}$)  corresponds to a spin gapped
phase with dominating SS, ES and CDW ordering. The corresponding correlations 
show a power-law decay at large distances:
\be \label{JxyJz_B_SS+ES}
\langle {\cal O}_{SS}(0){\cal O}_{SS}(r)\rangle =
\langle {\cal O}_{ES}(0){\cal O}_{ES}(r)\rangle 
  \simeq r^{-1},
\ee
\be \label{JxyJz_B_CDW}
\langle {\cal O}_{CDW}(0){\cal O}_{CDW}(r)\rangle \simeq  \cos(\pi r/2)\cdot
r^{-1} \, .
 \ee
In this case also the longitudinal ferromagnetic and the
$\eta$-superconducting correlations decay faster (\ref{FM-zz}) and
(\ref{eta}), while all other correlations are exponentially
suppressed.

In the sector {\bf C} ($J_{z}<- \left|J_{xy} \right|$) the gapless
Luttinger Liquid (LL) phase is realized. All correlations decay as
power-laws, however the transverse antiferromagnetic and triplet
superconducting instabilities are the dominating instabilities in this
sector. The corresponding correlations decay as
\bea 
\langle {\cal O}_{SDW^{xy}}(0){\cal O}_{SDW^{xy}}(r)\rangle 
&\simeq& \cos(\pi r/2) r^{-1-1/K_{s}^{\ast} }, \label{JxyJz_C_SDWxy}\\
\langle {\cal O}_{TS^{xy}}(0){\cal O}_{TS^{xy}}(r)\rangle 
&\simeq&  r^{-1-1/K_{s}^{\ast} }\, , \label{JxyJz_C_TSxy}
\eea
where the fixed point value of the spin LL parameter
\be\label{JxyJz_Ks}
K_{s}^{\ast} = 1 + \frac{1}{2\pi v_{F}}\sqrt{J_{z}^{2}-J_{xy}^{2}}>1\, .
\ee
Since the spin stiffness parameter $K_{s}^{\ast}>1$, the
longitudinal ferromagnetic and the $\eta$-superconducting correlations 
given by Eqs.~(\ref{JxyJz_A_TSz}) and (\ref{FM-zz}) decay faster and are the
subleading instabilities in this sector. The transverse
ferromagnetic correlations decay even faster as
\be 
\langle {\cal O}_{FM^{x,y}}(0){\cal
O}_{FM^{x,y}}(r)\rangle \simeq r^{-K_{s}^{\ast}-1/ K_{s}^{\ast}}\,
. \ee
However, within the accuracy of the used first order RG approach
their decay is the same as of the longitudinal ferromagnetic
correlations (\ref{FM-zz}).

Finally, the $CDW$, $SDW^{z}$ and the $TS^{0}$ correlations decay as 
\bea \label{FM and eta-SS}
\langle {\cal O}_{CDW}(0){\cal O}_{CDW}(r)\rangle &\simeq&
\langle {\cal O}^{z}_{SDW}(0){\cal O}^{z}_{SDW}(r)\rangle\nonumber\\
&\simeq & \cos(\pi r/2)\cdot r^{-1-K_{s}^{\ast}}\, ,\\
\langle {\cal O}_{TS^{0}}(0){\cal O}_{TS^{0}}(r)\rangle &\simeq & 
r^{-1-K_{s}^{\ast}}\, ,
\eea
and correspond to the weakest instabilities in this sector.

\begin{figure}[tbh]
\begin{center}
\includegraphics[width=0.4\textwidth]{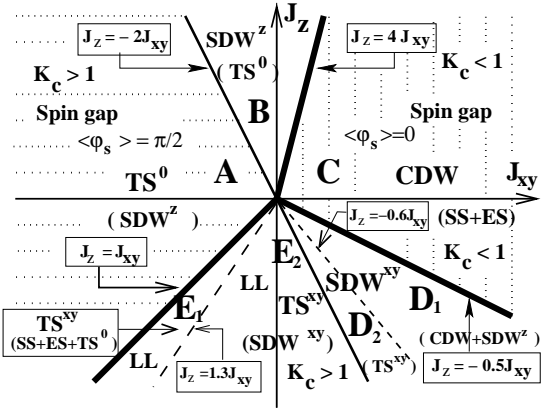}
\vspace{0.2cm}
\caption{The ground state phase diagram of the $1/3$-filled itinerant 
  $t-J_{xy}-J_{z}$ model. The thick solid lines denote lines in the
  parameter space where the spin gap opens and indicate borders
  between the weak-coupling limit phases.  The thin solid line
  corresponds to the line $K_{c}=1$ and marks the transition
  from the sector of phase diagram with dominating density
  instabilities ($K_{c}<1$) into the sector with dominating
  superconducting ordering ($K_{c}>1$). The dashed lines mark
  (qualitatively) the crossover areas between the Luttinger Liquid
  phases with different sets of subleading instabilities. The phases $A$ and
  $B$, as well as the $D_{2}$ and $E_{2}$ phases are characterized by
  an identical excitation spectrum and interchange of leading and
  subleading instabilities. The subleading instabilities are indicated
  in brackets.}
\label{Fig:Fig-2_PD_0333Filled_JxyJz}
\end{center}
\end{figure}

\subsubsection{$\nu \neq 1/4$}

At $\nu \neq 1/4$ the charge stiffness parameter $K_{c}\neq 1$ and therefore 
the line $J_{z}=-2J_{xy}$ divides the parameter space into two semiplanes: 
the part with dominating $CDW$ or $SDW$ instabilities at $K_{c}<1$ and the 
part with dominating superconducting instabilities at $K_{c}>1$. However, the 
effect of charge sector essentially depends on the band-filling. 

At $1/4 < \nu <1/2$, $K_{c} - 1 \sim {\mbox sign}(J_{z}+2J_{xy})$. 
Therefore the dynamical generation of a spin gap and
subsequent pinning of the spin field with vacuum expectation value 
$\langle \varphi_{s} \rangle =\pi/2 $ results to metallic phase with 
dominating Triplet Superconducting ($TS^{0}$) instability at 
$J_{xy} < 0$,\, $J_{xy} < J_{z}<-2J_{xy}$ and dominating antiferromagnetic
$SDW^{z}$ at $ -0.5J_{z} < J_{xy} < -0.5\cos(2\pi\nu) J_{z}$ 
(see Fig.~\ref{Fig:Fig-2_PD_0333Filled_JxyJz}, sectors $A$ and $B$ 
respectively). 

In the sector $C$ of the phase diagram, at
$-2J_{xy}\cos^{2}(\pi\nu) < J_{z} < -2J_{xy}/\cos(2\pi\nu)$, the spin gapped
metallic phase with dominating $CDW$ phase is realized.  

In the Luttinger Liquid sectors of the phase diagram D and E, at
$J_{z}<\min\{J_{xy},-2\cos^{2}(\pi\nu) J_{xy}\}$ the line
$J_{z}=-2J_{xy}$ marks the transition from a LL phase with dominating
$TS^{xy}$ instability at $J_{z}<\min\{J_{xy},-2J_{xy}\}$ (see
Fig.~\ref{Fig:Fig-2_PD_0333Filled_JxyJz}, sector $E$) to a LL phase
with dominating transverse antiferromagnetic instabilities $SDW^{xy}$
at $-2J_{xy}< J_{z}<-2\cos^{2}(\pi\nu)J_{xy}$ (see
Fig.~\ref{Fig:Fig-2_PD_0333Filled_JxyJz}, sector $D$). One has to note,
that for a given band-width $\nu \neq 1/4$ in the LL sectors D and E,
there are additional "subdominant order crossover" lines (marked by
dashed lines in Fig.~\ref{Fig:Fig-2_PD_0333Filled_JxyJz} and
\ref{Fig:Fig-3_PD_0167Filled_JxyJz}) which separate two areas with
two different subdominant order. In particular, in the subsector
$E_{1}$ the subdominant order is $SS+ES+TS^{0}$ and in the subsector
$E_{2}$ it is $SDW^{xy}$.  Similarly, in the subsector $D_{1}$ the
subdominant order is $CDW+SDW^{z}$ and in the subsector $D_{2}$ it is
$TS^{xy}$.

\begin{figure}[tbh]
\begin{center}
\includegraphics[width=0.4\textwidth]{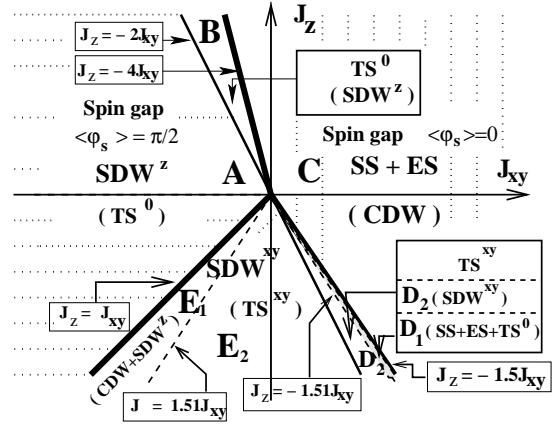}
\caption{The ground state phase diagram of the $1/6$-filled itinerant 
  $t-J_{xy}-J_{z}$ model.  Solid lines indicate borders between the
  weak-coupling limit phases. The thick solid lines denote lines in
  the parameter space where the spin gap opens and indicate borders
  between the weak-coupling limit phases.  The thin solid line
  $J_{z}=-2J_{xy}$ marks the transition from the sector of phase
  diagram with dominating density instabilities ($K_{c}<1$) into the
  sector with dominated superconducting ordering ($K_{c}>1$). The
  dashed lines mark (qualitatively) the crossover areas between the
  Luttinger Liquid phases with different sets of subleading
  instabilities. The phases $A$ and $B$, as well as the $D_{2}$ and $E_{2}$
  phases are characterized by an identical excitation spectrum and
  interchange of leading and subleading instabilities. The subleading
  instabilities are indicated in brackets.}
\label{Fig:Fig-3_PD_0167Filled_JxyJz}
\end{center} 
\end{figure}

On the other hand, at $\nu<1/4$ (see Fig.~\ref{Fig:Fig-3_PD_0167Filled_JxyJz})
the $A$ sector corresponds to the spin gapped metallic state with 
dominating $SDW^{z}$ instabilities, while triplet superconductivity 
$TS^{0}$ is the dominating instability in the narrow stripe of sector $B$. 
Similarly, in the sector $C$ at 
$J_{z}> \max\{-2J_{xy}/\cos(2\pi\nu),-2J_{xy}\cos^{2}(\pi\nu)\}$  
a spin gapped phase with dominating tendencies towards 
singlet superconducting ordering is realized. Similarly in the
Luttinger liquid phase dominant and subdominant instabilities in the
sectors $D$ and $E$ change places, i.e.\ in sector $D$ the dominant order
is $TS^{xy}$ and the subdominant order in sector $D_{1}$ is $SS+ES+TS^{0}$
and in sector $D_{2}$ it is $SDW^{xy}$. In sector $E$ the dominant order
is $SDW^{xy}$ and the subdominant order in sector $E_{1}$ is $SDW^{z}+CDW$
and in sector $E_{2}$ it is $TS^{xy}$.

\subsection{The $J_{z}=0$ case}

At $J_{z}=0$ the basic equations read:
\begin{eqnarray}\label{Kc-U-0a}
K_{c} \simeq  1 - \frac{1}{2\pi v_{F}}\left(U-J_{xy}\cos(2\pi\nu)\right)\, ,
\end{eqnarray}
and 
\bea
 2(K_{s}-1)  &\simeq& \frac{1}{\pi v_{F}}
\Big[U-J_{xy}\cos(2\pi\nu)\Big]\, ,
\label{Ks-U-0a}\\
2\pi m_{s} &\simeq& \frac{1}{\pi v_F}\Big[U-J_{xy} \Big]\, .
\label{Ms-U-0a}
\eea
Again we distinguish the cases $\nu = 1/4$ and $\nu \neq 1/4$
in the following.

\subsubsection{$\nu = 1/4$}
\label{sec-jz0-quarter}

At $U \neq 0$, even for $\nu = 1/4$ the charge stiffness parameter 
$K_{c}\neq 1$ and therefore the line $U=0$ divides the 
parameter space into two parts, the part with dominating $CDW$ or $SDW$
instabilities at $U>0$ ($K_{c}<1$) and the part with dominating 
superconducting instabilities at $U<0$ i.e. ($K_{c}>1$). 

\begin{figure}[tbh]
\begin{center}
\includegraphics[width=0.4\textwidth]{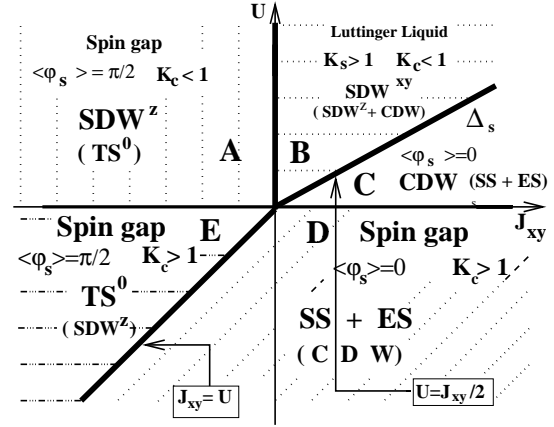}
\caption{The ground state phase diagram of the $1/4$-filled itinerant 
$t-J_{xy}-U$ model. Solid lines indicate borders between the weak-coupling 
limit phases. The thick solid lines denote lines in the 
parameter space where the spin gap opens and indicate borders between the 
weak-coupling limit phases.  The thin solid line $U=0$ marks the transition 
from the sector of phase diagram with dominating density instabilities 
($K_{c}<1$) into the sector with dominated superconducting ordering 
($K_{c}>1$). The subleading instabilities are indicated in brackets.}
\label{Fig:Fig-4_PD_025Filled_JxyU}
 \end{center}
 \end{figure}

At $J_{xy}<0$ and $U>J_{xy}$ the spin sector is gapped and the spin field 
gets ordered with vacuum expectation value $\langle\varphi_{s}\rangle
=\pi/2$. This leads to the suppression
of all instabilities whose power-law decay is less than $1/r^{2}$ 
except the $SDW^{z}$ and $TS^{0}$ ones, which show a power-law decay
\bea \label{UJxy_SDWz+TSz}
\langle {\cal O}_{SDW^{z}}(0){\cal O}_{SDW^{z}}(r)\rangle 
&\simeq& r^{-K_{c} }\cos(2\pi\nu r)\, ,\\
\langle {\cal O}_{TS^{0}}(0){\cal O}_{TS^{0}}(r)\rangle 
&\simeq & r^{-1/K_{c} }\, . 
\eea
Therefore, in the sector {\bf A}, at $J_{xy}<0<U$ the $SDW^{z}$ is the
dominating instability in the system and in the sector {\bf E} at 
$J_{xy}<U<0$ the triplet superconducting ordering dominates.

For $U<J_{xy}<0$ and for $U<J_{xy}/2$ the spin sector is also gapped but 
the vacuum expectation value of the ordered spin field $\langle\varphi_{s}
\rangle=0$. This leads to suppression of all instabilities with power-law 
decay less than $1/r^{2}$ 
except density-density and singlet superconducting instabilities,
which show the following power-law decay 
\bea \label{UJxy_CDW+SS+ES}
&\langle {\cal O}_{CDW}(0){\cal O}_{CDW}(r)\rangle  \simeq 
r^{-K_{c}} \cos(2\pi\nu r) \, ,&\\
&\langle {\cal O}_{SS}(0){\cal O}_{SS}(r)\rangle \simeq  
\langle {\cal O}_{ES}(0){\cal O}_{ES}(r)\rangle \simeq 
r^{-1/K_{c}}\, .& 
\eea
Therefore in the sector {\bf C} at $0<U<J_{xy}/2$ the CDW ordering 
dominates, while in the sector {\bf D} at $U < \min\{0, J_{xy}\}$ the 
singlet superconducting  order is realized. 

Finally, in the sector {\bf B} ($U>J_{xy}/2$) the LL phase with gapless 
charge and spin excitation spectrum and dominating easy-plane 
antiferromagnetic ordering is realized in the ground state. The corresponding 
correlations show a power-law decay
\be \label{UJxy_E_SDWz}
\langle {\cal O}_{SDW^{xy}}(0) {\cal O}_{SDW^{xy}}(r)\rangle 
\simeq  r^{-K_{c}-1/K_{s}^{\ast}}\, ,  
\ee
where the fixed point value of the spin stiffness parameter is given by
\be \label{UJxy_E_Ks}
K_{s}^{\ast}=1 + \frac{1}{2\pi v_F}\sqrt{J_{xy}(2U-J_{xy})}\, .
\ee

One can easily show that the $CDW$ and $SDW^{z}$ correlations decay faster
\bea \label{UJxy_E_CDW+SDWz}
\langle {\cal O}_{CDW}(0){\cal O}_{CDW}(r)\rangle &\simeq& \nonumber\\ 
\langle {\cal O}_{SDW^{z}}(0){\cal O}_{SDW^{z}}(r)\rangle &\simeq&
r^{-K_{c}-K_{s}^{\ast}}\, , 
\eea
and are the subleading instabilities in this sector.

\subsubsection{$\nu \not= 1/4$}

At $\nu \not= 1/4$ the phase diagram is qualitatively the same as 
at quarter-filling (see Fig.~\ref{Fig:Fig-4_PD_025Filled_JxyU}). The minor
difference consist of the $\nu$ dependence of the border lines which mark 
transitions between different phases. In particular: 

\begin{itemize}
\item[1)]{The semiplane in the parameter space corresponding to  
dominant density-density correlations (sectors A,B and C in the phase diagram) 
is separated from the semiplane  with dominant superconducting ordering 
(sectors E and D) by the line $U=J_{xy}\cos(2\pi\nu)$;}
\item[2)]{The line $U=J_{xy}\cos^{2}(\pi\nu)$ corresponds to the border line 
between the B and C sectors of the phase diagram.}
\end{itemize}

To conclude this section, we have shown that the weak-coupling phase
diagram of the non-half-filled itinerant $t-J-U$ model shows a triplet
superconducting ordering in the sector of the phase diagram with
dominating ferromagnetic exchange and easy-plane anisotropy.
However, as it was shown in Ref.~\cite{DJSZ_2004} in the case of
half-filled band and for transverse ferromagnetic exchange
stronger then some critical value $J_{xy}<J_{xy}^{c}$ ($ -J_{xy}^{c}
\simeq W$, where $W$ is the bandwidth), the transition into an
insulating phase with easy-plane type ferromagnetic ordering takes place.

Although the transition into the gapless phase with ferromagnetic order 
(ferrometal) at $\nu \neq 1/2$ is a pure finite-bandwith, strong coupling 
effect the very presence of this transition as well as a clear asymmetry 
between antiferromagnetic ($J_{xy}>0$) and ferromagnetic 
($J_{xy}<0$) exchange can be traced already within the weak-coupling 
treatment, if we take into 
account the Hartree regularization of the hopping amplitude given by 
$t_{\rm eff}=t \left(1 + \gamma J_{xy}/2\pi t \right)$. 
Here $\gamma(\nu)>0$ is a band-filling dependent parameter, which is of 
the order of unity for the band-fillings considered in this paper. 
It is clear that in the case of antiferromagnetic transverse exchange 
the effective bandwidth increases, while for 
ferromagnetic coupling it reduces. The weak-coupling approach fails when 
the effective dimensionless coupling constant 
$|g_{\it i}|= |{J_{xy}}|/{\pi v_F} $ becomes of the order of unity. 
Taking into account the Hartree renormalization of the Fermi velocity 
we estimate the range of applicability of the weak-coupling approach as 
$|J_{xy}|<\pi t$. In this sector of the parameter space with strong 
easy-plane anisotropy $|g_{\perp}|>|g_{0s}|$ the soliton mass of the 
sine-Gordon field (\ref{SGs}) is given by 
$M_{s} \simeq W\exp\left(-\pi W/2|J_{xy}|\right)$ \cite{GNT} and therefore 
for $|J_{xy}|\leq \pi t$ the spin gap increases with increasing strength 
of the transverse ferromagnetic exchange. However, since in the case of 
strong ferromagnetic exchange $J_{xy} \leq -\pi t$ the effective bandwidth 
$W \sim t_{\rm eff}$ tends to zero, the initial increase of the spin gap 
should change into a decrease caused by the collapse of the bandwidth $W$. 
As we show below, using the DMRG studies of chains up to $L=120$ sites, 
this is indeed the case. For 
$J_{xy}<0$, the spin gap as a function of the parameter $J_{xy}$ shows a 
bellshaped behavior with maximum at $J_{xy} \simeq -\pi t$ and reaches zero 
at $J_{xy} = J^{(c1)}_{xy} \simeq -4t$. At $J<J^{(c1)}_{xy}$ the ground state 
of the system is similar to that of the $t-J_{xy}$ model with gapless 
charge (due to the doping) and gapless spin (due to the in-plane, $XY$ 
character of the exchange) excitation spectrum.


\section{Numerical results for $J_z=0$ and $\nu=1/4$}

\label{sec:NumericalResults}

In order to check the validity of the picture suggested by the
bosonization results derived in the previous sections we use the 
density-matrix renormalization-group (DMRG) method 
\cite{White92,Peschel99,Schollwoeck05}. 
As in our previous study \cite{DJSZ_2004} it is applied to open chains 
up to $120$ sites keeping typically $400$ 
states in each block using the \emph{infinite-size} algorithm to 
determine the ground-state properties, including correlation functions.

Below we focus on the case of a quarter-filled band and $J_z=0$.
\begin{figure}[here]
\begin{center}
   \resizebox{0.9\columnwidth}{!}
 {\includegraphics{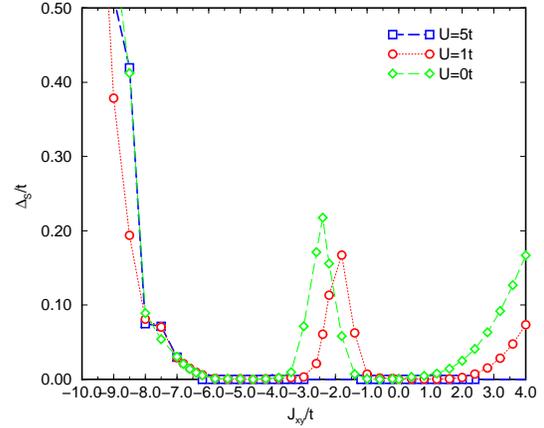}}
  \vspace{-0.0cm}
  \caption{Spin gap $\Delta_s$ of the itinerant $XY$ model at quarter filling, 
    $t=1$, $J_z=0$ and $U=0$ (diamonds), $U=1$ (circles) and $U=5$
    (squares).  The charge gap $\Delta_c$ vanishes for all $J_{xy}$.}
  \label{Fig:TRSC3_Fig5}
\end{center}     
\end{figure}

\subsection{Excitation spectrum}

First we determine the low-lying spin and charge excitations as
function of the transverse spin exchange $J_{xy}$. In the numerical
calculations, we have found that for open boundary conditions the
ground state energy for a system of $L$ sites with $N_{\uparrow}$
up-spin and $N_{\downarrow}$ down-spin electrons
$E_0^{(L)}(N_{\uparrow},N_{\downarrow})$ remains in the sector with
the "z" component of the total spin $S^{z}_{tot}=0$ (i.e.
$N_{\uparrow} = N_{\downarrow} =N/2 $, where
$N=N_{\uparrow}+N_{\downarrow}$ is the total number of electrons) for
all parameter values studied here.

As it is commonly used in literature, gaps to excitations classified
as {\it charge excitations} are calculated by taking the difference
between ground-state energies with different number of particles. It
is convenient to stay in the sector with $S^{z}_{tot}=0$ and therefore
the charge gap is evaluated by
\bea
\Delta_c(L) &=& \frac{1}{2}
  \left[E_0^{(L)}\left(\frac{N}{2}+1,\frac{N}{2}+1\right)\right.\nonumber\\
  &&\hspace{-0.8cm}+E_0^{(L)}\left(\frac{N}{2}-1,\frac{N}{2}-1\right)
\left.-2E_0^{(L)}\left(\frac{N}{2},\frac{N}{2}
     \right)\right]\, .
\eea

We determine the spin gap as the difference between the lowest energy
in the sector with $S^{z}_{tot}=0$ ($N_{\uparrow} - N_{\downarrow}
=2$) and the ground state energy
\begin{eqnarray}
  \Delta_s(L) &=& E_0^{(L)}\left(\frac{N}{2}+1,\frac{N}{2}-1\right)-
  E_0^{(L)}\left(\frac{N}{2},\frac{N}{2}\right).
  \label{eq:spin_charge_gap_def}
\end{eqnarray} 

The results for finite chains are extrapolated for $L\to\infty$ by
fitting a polynomial in $1/L$. As expected the charge gap 
$\Delta_c=\lim_{L\to\infty}\Delta_c(L)$ vanishes
for all values of $J_{xy}$. On the other hand, the spin gap 
$\Delta_s=\lim_{L\to\infty}\Delta_s(L)$ shows a nontrivial behavior where
five different regimes can be distinguished 
(Fig.~\ref{Fig:TRSC3_Fig5}).
\begin{figure}[here]
\begin{center}
  \resizebox{0.95\columnwidth}{!}
  {\includegraphics{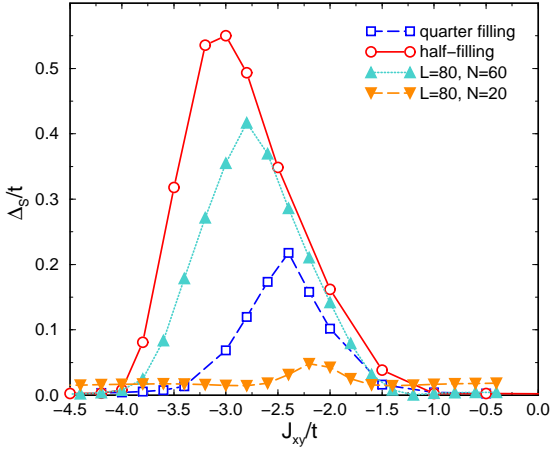}}
  \vspace{-0.0cm}
  \caption{Spin gap $\Delta_s$ of the itinerant ferromagnetic $XY$ model 
   ($J_{xy}<0$) at $U=0$ and for fillings $\nu=1/2,1/4,3/8,1/8$.}
\label{Fig:spectrum_t1_Jxy_U0.eps} 
\end{center}     
\end{figure}
In agreement with predictions of the weak-coupling treatment, at $U=0$
and weak exchange $\left|J_{xy}\right|$ the spin sector is gapped for
both signs of the transverse exchange. With increasing antiferromagnetic 
exchange the spectrum remains gapped, while in the case of ferromagnetic 
exchange, at $J_{xy} \simeq -2.5t$ the spin gap starts to decrease and becomes 
zero at $J_{xy}^{c1} \simeq -3.4t$. The spin gapless phase remains till 
$J_{xy}^{c2} \simeq -5.8t$ where the spin gap opens once again. 

\begin{figure}[t]
\ \phantom{as}\\
\begin{center} 
\resizebox{0.9\columnwidth}{!}
{\includegraphics{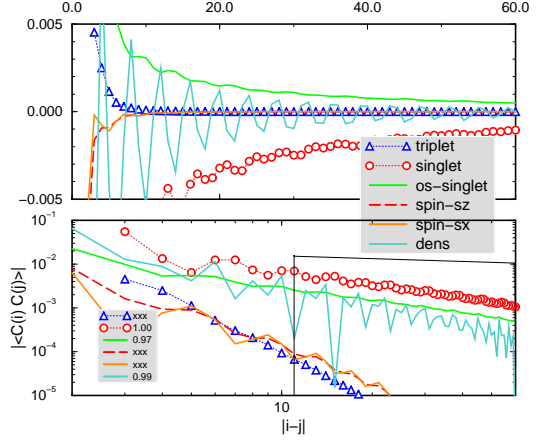}}
\caption{Correlation functions, plotted against the real space distance
      $|i-j|$, in the case of strong antiferromagnetic exchange
      $J_{xy}=3.8t$ and $U=0$. The lower figure shows the
      decay of the correlations, plotted on a double logarithmic scale.   
      Numbers in the inset of the lower figure are the exponents. 
      The notation "xxx" corresponds to the case of exponentially decaying 
      correlations.}
\label{Fig:qf_pair_t1_Jxy3o8_U0.eps}  
\end{center}    
\end{figure}
Thus three gapful regimes (for large ferro- and antiferromagnetic 
couplings and for intermediate ferromagnetic exchange) are separated by 
two gapless regimes. As it is seen from  
Fig.~\ref{Fig:TRSC3_Fig5} a repulsive Coulomb interaction 
$U>0$ only changes this behaviour quantitatively. The gap is slightly
suppressed and the gapful regimes appear at larger values of the exchange 
interaction $J_{xy}$.

We have also checked the band-filling dependence of the spin-gapped
phase for intermediate feromagnetic exchange. As it follows from 
Fig.~\ref{Fig:spectrum_t1_Jxy_U0.eps} the dome-type dependence of the
spin gap on the increasing ferromagnetic transverse qualitative
remains unchanged at $\nu \neq 1/2$. Although position and value
of the maximum of the spin gap depends on the band filling 
(Fig.~\ref{Fig:spectrum_t1_Jxy_U0.eps}) the effect
of the closing of the spin gap at large ferromagnetic exchange is 
band-filling independent.

\subsection{Correlation functions at $U=0$}

To investigate the nature of ordering in the different phases we study the 
behavior of the correlation functions. In the sectors with gapless excitation 
spectrum we expect the usual expression for correlation functions
\begin{equation}
  C(r)\equiv
  \langle{\cal O}^{\dagger}(r){\cal O}^{\vphantom{\dagger}}(0)\rangle \sim
  A_1r^{-\theta_1} \,+ \, \cos\left(2k_{F}r\right)A_{2}r^{-\theta_2}
\end{equation}
consisting of a smooth part decaying with exponent $\theta_1$
and an oscillating part decaying with $\theta_2$.
In determining the asymptotics of correlation functions we focus on the 
dominating part given by $\theta=\min\{\theta_1,\theta_2\}$.

In the following we will present results for correlation functions in
different sectors of the phase diagram.

\subsubsection{\bf Sector \textrm{I}: $J_{xy}>0$.\\
\hspace{6mm} The SS $+$ ES $+$ CDW phase.}

We start our consideration from the case antiferromagnetic exchange. In 
Fig.~\ref{Fig:qf_pair_t1_Jxy3o8_U0.eps} we have plotted results of DMRG
calculations for correlation functions at $J_{xy}=3.8t$ and $U=0$.  
\begin{figure}[t]
\ \phantom{as}\\
\begin{center} 
\resizebox{0.9\columnwidth}{!}
{\includegraphics{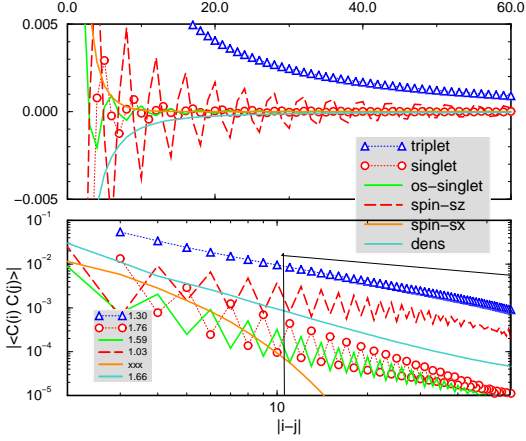}}
  \caption{Correlation functions, plotted against the real space distance
    $|i-j|$ in the case of ferromagnetic exchange
    $J_{xy}=-2.8t$ and $U=0$. The lower figure shows the 
    decay of the correlations, plotted on a double logarithmic scale.
    Numbers in the inset of the lower figure are the exponents. 
    The notation "xxx" corresponds to the case of exponentially decaying 
    correlations.}
  \label{Fig:qf_pair_t1_Jxym2o8_U0.eps}  
\end{center}    
\end{figure}

As it is clearly seen from Fig.~\ref{Fig:qf_pair_t1_Jxy3o8_U0.eps}, in
an excellent agreement with the bosonization results, the triplet
superconducting and antiferromagnetic correlations are completely
suppressed in this case, while the singlet superconducting and
density-density correlations show an identical power-law decay with
critical indices equal to one. The very small deviation of the
numerically evaluated values of the exponents $\theta_{ES}=0.97$ and
$\theta_{CDW}=0.98$ from the analytically predicted value
$\theta_{SS}=\theta_{ES}=\theta_{CDW}=1$ reflects the high accuracy of
the obtained numerical results in this sector of the phase diagram.

\subsubsection{\bf Sector \textrm{II}: $J_{xy}^{c1}<J_{xy}<0$.\\ 
\hspace{6mm}The SDW$^{z} +$  TS$^{0}$ phase}

Let us now consider the case of ferromagnetic exchange. At $J_{xy}<0$
and $U=0$.  the weak-coupling bosonization results predict exponential
suppression of the CDW and singlet correlations, whereas SDW$^{z}$ and
triplet correlators TS$^{0}$ show a power-law decay (cf. with
Eq.~(\ref{JxyJz_A_SDWz})-(\ref{JxyJz_A_TSz})).  Furthermore, they are
the dominating instabilities in this phase.

Fig.~\ref{Fig:qf_pair_t1_Jxym2o8_U0.eps} displays DMRG results for the 
correlation functions. One can clearly observe a strong SDW$^{z}$ and TS$^{0}$ 
correlation in the ground state. In addition, from the double logarithmic plot 
one obtains that the in-plane magnetic correlations decay 
exponentially, while the density and singlet-superconducting correlations show
an almost identical power-law decay at large distances. This is expected from
the bosonization results, since, due to the pinning of the spin Bose field 
with vacuum expectation value $\la \sqrt{2 \pi K_{s}}\varphi_{s}\ra = \pi/2$,
the in-plane magnetic correlations which are determined by the dual-field 
decay exponentially. The same reason of spin-field pinning leads to
suppression of the oscillating part in the density correlations (clearly seen
in Fig.~\ref{Fig:qf_pair_t1_Jxym2o8_U0.eps}) and to the suppression of the 
smooth part of the on-site singlet correlations. The latter is the
reason why the on-site and extended singlet-pair correlations show an 
identical behavior.

\subsubsection{\bf Sector \textrm{III}: $J_{xy}^{c2}<J_{xy}<J_{xy}^{c1}$.\\ 
\hspace{6mm} The ferrometallic phase}

In Fig.~\ref{Fig:qf_pair_t1_Jxym5o0_U0.eps} we have shown results of
DMRG calculations for correlation functions for strong ferromagnetic
coupling $J_{xy}=-5.0t$. In this sector, charge and 
spin gap vanish and the system shows properties of a metal with
gapless spin degrees of freedom. As it is clearly seen, the transverse 
ferromagnetic and triplet superconducting instabilities dominate in the ground
state. The lower figure in Fig.~\ref{Fig:qf_pair_t1_Jxym5o0_U0.eps}
displays correlation functions plotted in a double logarithmic scale.
As it follows from this figure, despite the gapless character of spin
excitations, the singlet superconducting correlations are suppressed
exponentially.
\begin{figure}[here]
\ \phantom{as}\\
\begin{center} 
\resizebox{0.9\columnwidth}{!}
{\includegraphics{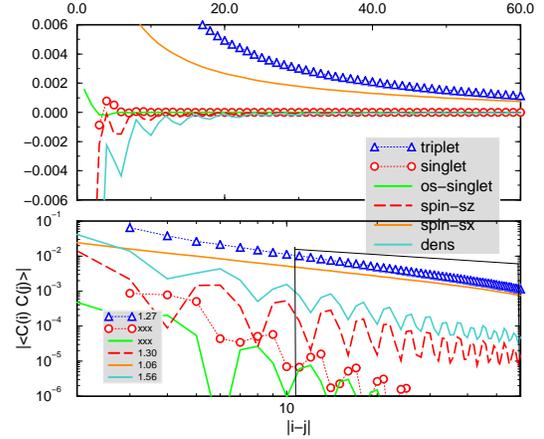}}
\caption{Correlation functions, plotted against the real space distance
      $|i-j|$, in the case of strong antiferromagnetic exchange
      $J_{xy}=-5.0t$ and $U=0$. The lower figure shows 
      decay of the correlations, plotted on a double logarithmic scale.
      Numbers in the inset of the lower figure are the exponents. 
      The notation "xxx" corresponds to the case of exponentially decaying 
      correlations.}
\label{Fig:qf_pair_t1_Jxym5o0_U0.eps}  
\end{center}    
\end{figure}
At the same time, the SDW$^{z}$ and CDW correlations also show
power-law behavior, but decay slightly faster then the in-plane spin
correlations. Clearly in this sector the system shows very
unconventional behaviour, not consistent with the standard
weak-coupling results.

\subsubsection{\bf Sector \textrm{IV}: $J_{xy}<J_{xy}^{c2}$.\\ 
\hspace{6mm}The triplet superconducting phase.}

This is the most intriguing sector of the phase diagram. The spin excitation
spectrum is gapped. Moreover, as it is clearly seen from the 
Fig.~\ref{Fig:TRSC3_Fig5}, the opening of the spin gap at
$J_{xy}<J_{xy}^{c2}\simeq -5.8t$ is {\em independent} of the on-site
repulsion $U$. This indicates that in this sector of the phase diagram the
itinerant nature of the electron system is completely lost and the phase
diagram is completely determined by the properties of the effective
ferromagnetic $t-J_{xy}$ model. 
We expect that this sector of the phase diagram 
corresponds to a triplet superconducting phase.

\section{Conclusions}
\label{sec-conclusion}

Motivated by recent experimental findings that show evidence for the 
competition or even coexistence of superconductivity and magnetism we 
have continued our studies of the ground state properties of an itinerant 
$XY$ model. Using a composite approach based on weak-coupling
bosonization and DMRG studies for chains up to $L=120$ sites we have
studied the ground state phase diagram of the itinerant $XY$ model 
away from half-filling. 
\begin{figure}[here]
\begin{center} 
\resizebox{0.9\columnwidth}{!}
{\includegraphics{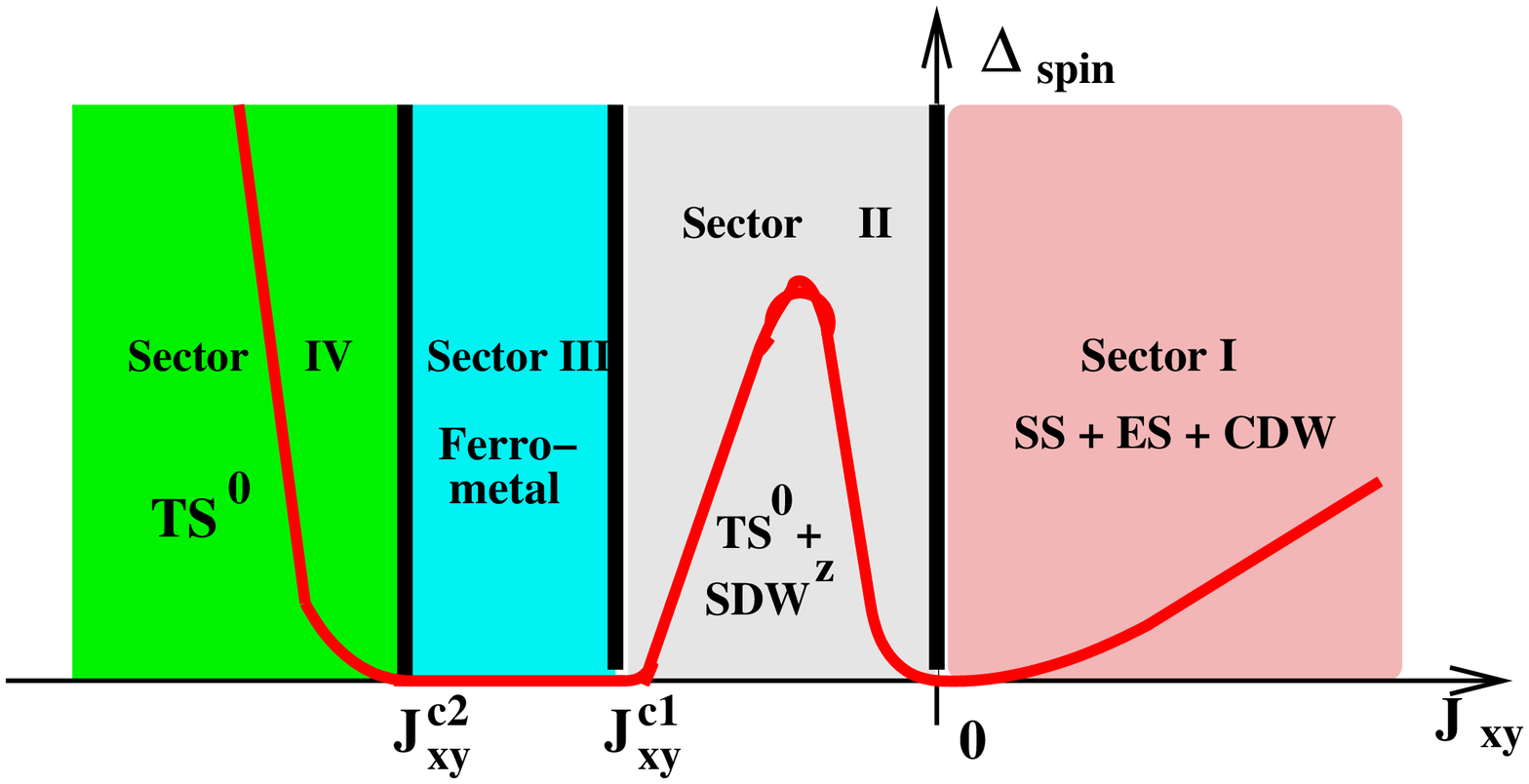}}
\caption{Qualitative form of the ground state phase diagram for the
$t-J_{XY}-U$ model at quarter-filling and $t=1$ and $U>0$. 
The bold red line shows the spin gap as a function of the parameter $J_{xy}$.}
\label{Fig:TRSC3_Fig_10}
\end{center}    
\end{figure}

Depending on the model parameters $J_{xy}$, $U\geq 0$ and the band-filling
$\nu$ we have found evidence for five different phases in the ground
state.  Fig.~\ref{Fig:TRSC3_Fig_10} summarizes our findings for
the particular case of a quarter-filled band. Within the considered
range of parameters the charge gap is always zero. The behavior of the
spin gap as function of the spin-coupling $J_{xy}$ allows to
distinguish the following different phases:
\begin{itemize}
\item[1.]{For antiferromagnetic interactions $J_{xy} >0$ the spin
    gapful metallic phase with dominating singlet superconducting and
    density-density correlations.}
\item[2.]{Approaching the line $J_{xy} =0$ the spin gap closes and at
    this point the system shows properties of a Luttinger metal.}
\item[3.] In the ferromagnetic sector at $-3.8t=J_{xy}^{c1}<J_{xy}<0$
  the spin gap is finite. It opens for arbitrary weak ferromagnetic
  $XY$-exchange, shows a dome-type shape and closes at
  $J_{xy}=J_{xy}^{c1}$. In this sector of the phase diagram the system
  the system displays properties of a spin gapped metal with
  coexistence of triplet superconducting and spin-density-wave
  (SDW$^{z}$) instabilities.
\item[4.] At $-5.8t=J^{c2}_{xy}<J_{xy}<J^{c1}_{xy}$ a ferrometallic
  phase with gapless spin excitations and strongly dominating triplet
  superconducting and transverse ferromagnetic instabilities is
  realized in the ground state.
\item[5.] At $J_{xy}<J^{c2}_{xy}$ the spin gap opens once again. We
  expect that in this phase the system shows properties of a triplet
  superconductor. We have to stress that the effective ferromagnetic
  $t-J_{xy}$ model, which governs the behaviour of the system in this
  sector of the phase diagram, requires very detailed separate studies.
\end{itemize}

Although the presented numerical results are restricted to the case of
quarter-filled band, our analysis indicates that qualitatively the
phase diagram remains similar at $\nu \neq 1/4$. However, deviations
from the commensurate value of the quarter-filled band remove the
degeneracy in favour of density-density type ordering at
$J_{xy}\cos(2\pi\nu)<0 $ and superconducting type ordering at
$J_{xy}\cos(2\pi\nu)>0$.

One of the interesting perspective for future studies would be
the investigation of a genuine ferromagnetic $t-J$-type model which should
help to understand the limit of strong Coulomb repulsion better.

\paragraph{Acknowledgments}

This work has been performed within the research program of
the SFB 608 funded by the DFG. We thank Erwin M\"uller-Hartmann, 
Achim Rosch, Didier Poilblanc and Matthias Vojta for interesting discussions. 
IT also acknowledges support from the DAAD scholarship programs.



\begin{thebibliography}{99}

\bibitem{dagotto05}{E. Dagotto, Science {\bf 309}, 257 (2005)}

\bibitem{Chubukov_Pines_book_2003}{A. V. Chubukov, D. Pines and J. Schmalian 
  {\em in "The Physics of Superconductors"} edited by K.~H. Bennemann and 
  J.~B. Ketterson, Springer-Verlag 2003, p. 495.}

\bibitem{Manske_book_2004}{ D. Manske, {\em Theory of Unconventional
      Superconductors} (Springer, Heidelberg, 2004).}

\bibitem{MacKenzieMaeno}{For a review, see A.~P. Mackenzie and Y. Maeno,
    Rev. Mod. Phys. \textbf{75}, 657 (2003).}

\bibitem{UGE2}{S.~S. Saxena, P. Agarwal, K. Ahilan, F.~M. Grosche,
    R.~K.~W. Haselwimmer, M.~J. Steiner, E. Pugh, I.~R. Walker,
    S.~R. Julian, P. Monthoux, G.~G. Lonzarich, A. Huxley, I. Sheikin,
    D. Braithwaite and J. Flouquet, Nature \textbf{406}, 587 (2000).}

\bibitem{URhGe}{D. Aoki, A. Huxley, E. Ressouche, D. Braithwaite,
    J. Flouquet, J-P. Brison, E. Lhotel and C. Paulsen,
    Nature \textbf{413}, 613 (2001).}

\bibitem{ZrZn2}{C. Pfleiderer, M. Uhlarz, S.~M. Hayden, R. Vollmer,
    H.~v.~L\"{o}hneysen, N.~R. Bernhoeft and G.~G. Lonzarich,
    Nature \textbf{412}, 58 (2001).}

\bibitem{RS}{T.~M. Rice and M. Sigrist, J. Phys. Condens. Matter \textbf{7}, 
    L643 (1995).}

\bibitem{MS}{I.~I. Mazin and D.~J. Singh, Phys. Rev. Lett. \textbf{79}, 
    733 (1997).} 

\bibitem{Sigrist}{M. Sigrist, D. Agterberg, A. Furusaki, 
    C. Honerkamp, K.~K. Ng, T.~M. Rice and M.~E. Zhitomirsky, 
    Physica C \textbf{317-318}, 134 (1999).}

\bibitem{Kirpatrick1}{T.~R. Kirkpatrick, D. Belitz, T. Vojta and R. Narayanan,
    Phys. Rev. Lett. \textbf{87}, 127003 (2001).}

\bibitem{Singh_Mazin_2002}{D.~J. Singh and I.~I. Mazin, 
    Phys. Rev. Lett. \textbf{88}, 187004 (2002).}

\bibitem{WalkerSamokhin}{M.~B. Walker and K.~V. Samokhin, 
    Phys. Rev. Lett. \textbf{88}, 227001 (2002).}

\bibitem{Chubukov2}{A.~V. Chubukov, A.~M. Finkel'stein, R. Haslinger and 
    D.~K. Morr, Phys. Rev. Lett. \textbf{90}, 077002 (2003).}

\bibitem{Buzdin}{A.~I. Buzdin, A.~S. Mel'nikov, Phys. Rev. B \textbf{67}, 
    020503 (2003).}

\bibitem{Kirpatrick2}{T.~R. Kirkpatrick and D. Belitz, 
    Phys. Rev. B \textbf{67}, 024515 (2003).}

\bibitem{Mazin_Singh_2004}{I.~I. Mazin and  D.~J. Singh, Phys. Rev. B 
\textbf{69}, 020402 (2004).}

\bibitem{Mazin_2004} {M.~D. Johannes, I.~I. Mazin, D.~J. Singh, and 
D.~A. Papaconstantopoulos, Phys. Rev. Lett. \textbf{93}, 097005 (2004).}

\bibitem{Ishiguro}{T. Ishiguro, K. Tamaji and G. Saito,
    {\em Organic Superconductors} 2nd ed. (Springer 1998).}

\bibitem{Jerome_1980}{D. J\'erome, A. Mazaud, M. Ribault and K. Bechgaard,
    J. Phys. (Paris) Lett. \textbf{41}, L95 (1980).}

\bibitem{Jerome_1991}{D. J\'erome, Science \textbf{252}, 1509 (1991).} 

\bibitem{TrSc_BechgaardSalts_Exp}{I.~J. Lee, M.~J. Naughton, 
    G.~M. Danner and P.~M. Chaikin, Phys. Rev. Lett. \textbf{78}, 3555 (1997); 
    S. Belin and K. Behnia, Phys. Rev. Lett. \textbf{79}, 2125 (1997); 
    I.~J. Lee, S.~E. Brown, W.~G. Klark, M.~J. Strouse, M.~J. Naughton, 
    W. Kang and P.~M. Chaikin, Phys. Rev. Lett. \textbf{88}, 017004 (2002); 
    I.~J. Lee, D.~S. Show, W.~G. Klark, M.~J. Strouse, M.~J. Naughton, 
    P.~M. Chaikin and S.~E. Brown, Phys. Rev. B. \textbf{68}, 092510 (2003).}


\bibitem{Lebed_2000} A.~G. Lebed, K.~Machida, and M.~Ozaki, Phys. Rev. B 
\textbf{62}, R795 (2000).

\bibitem{Kuroki_2001} K.~Kuroki, R.~Arita, and H~Aoki, Phys. Rev. B 
\textbf{63}, 094509 (2001). 

\bibitem{Siginishi_2004} Y. Suginishi and H. Shimahara, J. Phys. Soc. Jpn. 
\textbf{73}, 3121 (2004).  

\bibitem{Tanaka_Kuroki_2004} Y. Tanaka and K. Kuroki, Phys. Rev. B 
\textbf{70}, 060502 (R) (2004). 

\bibitem{Kuroki_et_al_2004} K.~Kuroki, Y~Tanaka, T.~Kimura, and R.~Arita,
Phys. Rev. B \textbf{69}, 214511 (2004).  

\bibitem{FS_CM_0411013} Y. Fuseya and Y. Suzumura, cond-mat/0411013.  

\bibitem{Nishimoto_et_al_2005}Y.~Ohta, S.~Nishimoto, T.~Shirakawa, and
Y.~Yamaguchi cond-mat/0504191.

\bibitem{JM_2000}{G.~I. Japaridze and E. M\"uller-Hartmann, 
    Phys. Rev. B \textbf{61}, 9019 (2000).}

\bibitem{DJSZ_2004} C. Dziurzik, G.I. Japaridze, A. Schadschneider, and 
J. Zittartz, Eur. Phys. J. B {\bf 37}, 453 (2004).

\bibitem{GNT} A.~O. Gogolin, A.~A. Nersesyan and A.~M. Tsvelik,
              {\em Bosonization and strongly correlated systems},
              Cambridge University Press (1998).


\bibitem{KT}{J.~M. Kosterlitz and D.~J. Thouless, 
             J. Phys. C {\bf 11}, 1583 (1973).}

\bibitem{Wiegmann} {P. Wiegmann, J. Phys. C {\bf 11}, 1583 (1978).}

\bibitem{ME}{K.~A. Muttalib and V.~J. Emery, Phys. Rev. Lett. {\bf 57}, 
    1370 (1986); T. Giamarchi and H.~J. Schulz, Jour. Phys. (Paris) {\bf 49}, 
    819 (1988); Phys. Rev. B {\bf 33}, 2066 (1986).}

\bibitem{Yang}{C.~N. Yang, Phys. Rev. Lett. {\bf 63}, 2144 (1989).}

\bibitem{White92}{S.~R. White, Phys. Rev. Lett. {\bf 69}, 2863 (1992); 
    Phys. Rev. B {\bf 48}, 10345 (1993).}

\bibitem{Peschel99}{\textit{Density-Matrix Renormalization}, edited by  
    I. Peschel, X. Wang, M. Kaulke and K. Hallberg (Springer, 1999).}

\bibitem{Schollwoeck05}{U. Schollw\"ock, Rev.\ Mod.\ Phys.\ {\bf 77}, 259
(2005).}

\end{thebibliography}
\end{document}